\documentclass[fleqn,usenatbib]{mnras}

\usepackage{newtxtext,newtxmath}

\usepackage[T1]{fontenc}

\DeclareRobustCommand{\VAN}[3]{#2}
\let\VANthebibliography\thebibliography
\def\thebibliography{\DeclareRobustCommand{\VAN}[3]{##3}\VANthebibliography}

\usepackage{graphicx}	
\usepackage{amsmath}	

\usepackage[flushleft]{threeparttable}
\usepackage{ulem}
\usepackage{multirow}

\def\h{$^{\rm h}$}
\def\m{$^{\rm m}$}

\def\fss{\hbox{$.\!\!^{\rm s}$}}        
\def\degs{\ifmmode ^{\circ}\else$^{\circ}$\fi}
\def\amin{\ifmmode ^{\prime}\else$^{\prime}$\fi}
\def\asec{\ifmmode ^{\prime\prime}\else$^{\prime\prime}$\fi}
\def\farcs{\hbox{$.\!\!^{\prime\prime}$}}  

\def\sw{\textit{Swift}}
\def\fermi{\textit{Fermi}}
\def\eros{\textit{eROSITA}}
\def\gaia{\textit{Gaia}}
\def\ps{Pan-STARRS}
\def\gem{\textit{Gemini}}

\def\msun{M$_{\odot}$}
\def\rsun{R$_{\odot}$}

\def\nh{$N_{\rm H}$}
\def\ergs{erg~s$^{-1}$}
\def\flux{erg~s$^{-1}$~cm$^{-2}$}

\def\masyr{mas~yr$^{-1}$}

\def\fgl{4FGL J2249.4$+$6229}
\def\src{J2249}
\def\lsxps{LSXPS~J224828.5$+$622211}
\def\srge{SRGe~J224828.4$+$622210}

\title[A new redback candidate \fgl]{
Nature of 4FGL J2249.4+6229: Evidence for a redback system with a cool companion and low X-ray and $\gamma$-ray luminosities}

\author[A. V. Karpova et al.]{
A. V. Karpova,$^{1}$
D. A. Zyuzin,$^{1}$
S. V. Zharikov$^{2}$\thanks{E-mail: zhar@astro.unam.mx (SVZ)}
and M. R. Gilfanov$^{3,4}$
\\
$^{1}$Ioffe Institute, Politekhnicheskaya 26, St. Petersburg, 194021, Russia \\
$^{2}$Instituto de Astronom\'ia, Universidad Nacional Aut\'onoma de M\'exico, Apdo. Postal 106, Baja California, M\'exico, 22860\\
$^{3}$Space Research Institute of the Russian Academy of Sciences, Profsoyuznaya 84/32, 117997 Moscow, Russia \\
$^{4}$Max-Planck-Institut f\"ur Astrophysik, Karl-Schwarzschild-Str. 1, D-85741 Garching, Germany
}

\date{Accepted XXX. Received YYY; in original form ZZZ}

\pubyear{\the\year{}}

\begin{document}
\label{firstpage}
\pagerange{\pageref{firstpage}--\pageref{lastpage}}
\maketitle

\begin{abstract}
We report the identification of the likely X-ray and optical counterpart to the unassociated \fermi\ source \fgl.
To clarify its nature, we investigate the X-ray data from \sw/XRT and SRG/\eros\ as well as photometric data from optical catalogues and archival spectroscopic data from the \gem-North telescope.
Using Zwicky Transient Facility data spanning over 6.6 yr, we confirmed a period of $\approx$5.6 h likely associated with the orbital motion in a binary system.
The folded light curves have a smooth sinusoidal shape with two peaks per period and the amplitude of $\approx$0.2 mag. 
The X-ray spectra of the source are well fitted by an absorbed power law with the photon index of $\approx$2.0 and unabsorbed flux of $\approx$1.4$\times10^{-13}$ \flux.
All these together with the X-ray to optical flux ratio of $\sim$0.2 implies that \fgl\ is a promising redback candidate.
Fitting the optical light curves with the direct heating model, we obtained the companion mass of $\approx$0.5~\msun\ and temperature of $\approx$3600 K implying an M-type star. 
This places it among the coldest and most massive companions known in redback systems.
Optical spectra confirms the M-type star and shows the broad asymmetric H$\alpha$ emission line.
For the distance of 500--550 pc derived from the optical data, the source can be the redback with the lowest X-ray and $\gamma$-ray luminosities. 
\end{abstract}
\begin{keywords}
binaries: close -- stars: individual: \fgl\ -- stars: neutron -- X-rays: binaries
\end{keywords}


\section{Introduction}
\label{sec:intro}

Thanks to the \fermi\ Gamma-ray Space Telescope, a large number of binary millisecond
pulsars (MSPs) have been discovered in recent years. Among them, there are members of
the `spider' family: redbacks (RBs) and black widows (BWs) \citep{roberts2013}. 
They are characterised by short orbital periods, $P_b < 1$~d, and low-mass companion
stars. Companions of BWs are semi-degenerate and have masses $M_{\rm c}<0.05$~\msun, while 
RBs possess non-degenerate and more massive ($M_{\rm c}\approx$~0.1--1~\msun) secondaries.
Three RBs -- the so-called transitional millisecond pulsars (tMSPs) -- show transitions between rotation-powered pulsar and active X-ray states 
\citep{papitto&demartino2022} confirming the evolutionary link between low-mass X-ray 
binaries (LMXBs) and MSPs. The pulsar mass in spider systems can exceed 2~\msun\ which
makes their studies particularly important for constraining the equation of state of 
the superdense matter in neutron stars (NSs) interiors \citep[e.g.][]{kumar2023}.

Spiders show orbital modulation in the optical due to the heating of the companion 
by the pulsar and/or its ellipsoidal shape \citep[e.g.][]{Callanan1995, draghis2019, clark2021, matasanchez2023, Kandel2023, dodge2024}, and in X-rays due to the intrabinary shock 
(IBS) produced by the interacting winds from the pulsar and its companion 
\citep[e.g.][]{Romani&Sanchez2016, sullivan&romani2024}. 
The optical spectra are dominated by the cool secondary star \citep[e.g.][]{vanKerkwijk2011, Romani&Shaw2011, romani2015, strader2019, Turchetta2025}, while the X-ray spectra are usually non-thermal and attributed to the IBS emission \citep[e.g.][]{Kulkarni1992, Arons1993, kandel2019, sullivan&romani2025, Sullivan2026}. However, a thermal component originating from the pulsar polar 
caps can also be present.

To date, about 80 confirmed spider pulsars have been detected in the Galactic 
field \citep{spidercat}. However, a significant challenge in detecting new spider 
pulsars via radio surveys is the periodic obscuration of the pulsar's radio emission 
by ablated material from the companion star, which prevents the detection of 
pulsations.  Nevertheless, these systems can be identified through multi-wavelength
investigations, especially in the optical and X-rays, of likely counterparts to 
unassociated \fermi\ sources \citep[e.g.][]{salvetti2017, braglia2020}. Currently,
such studies have revealed about 30 spider candidates 
\citep[e.g.][]{li2021,swihart2021,swihart2022,halpern2022,karpova2023,karpova2025,zyuzin2024}.

Here we report on the discovery of the likely X-ray and optical counterpart 
to the unassociated $\gamma$-ray source \fgl\ (hereafter \src). 
According to the \fermi\ Large Area Telescope 14-Year Point Source Catalog (4FGL-DR4; \citealt{4fgl-dr4}), its flux in the 0.1--100 GeV band is $F_\gamma \approx (9.98 \pm 1.84) \times 10^{-12}$ \flux.
In addition, its steady emission and a spectrum, which can be described by the LogParabola model, suggest it could be a pulsar.
Indeed, \citet{mayer&becker2024} provided the probability of 0.92 for the source 
to be a pulsar rather than a blazar.
We find that the properties of the \src\ presumed counterpart are typical for RB 
systems. 

The X-ray and optical identification of the source is described in Sec.~\ref{sec:data}. Sections~\ref{sec:lc} and \ref{sec:x-rays} contain the analysis of the data. The results and conclusions are given in Secs.~\ref{sec:discussion} and \ref{sec:conclusions}.


\section{X-ray/optical counterpart identification}
\label{sec:data}

\begin{figure}
\begin{center}
\includegraphics[width=0.8\linewidth, trim={0.6cm 0.5cm 0.3cm 0.9cm}, clip]{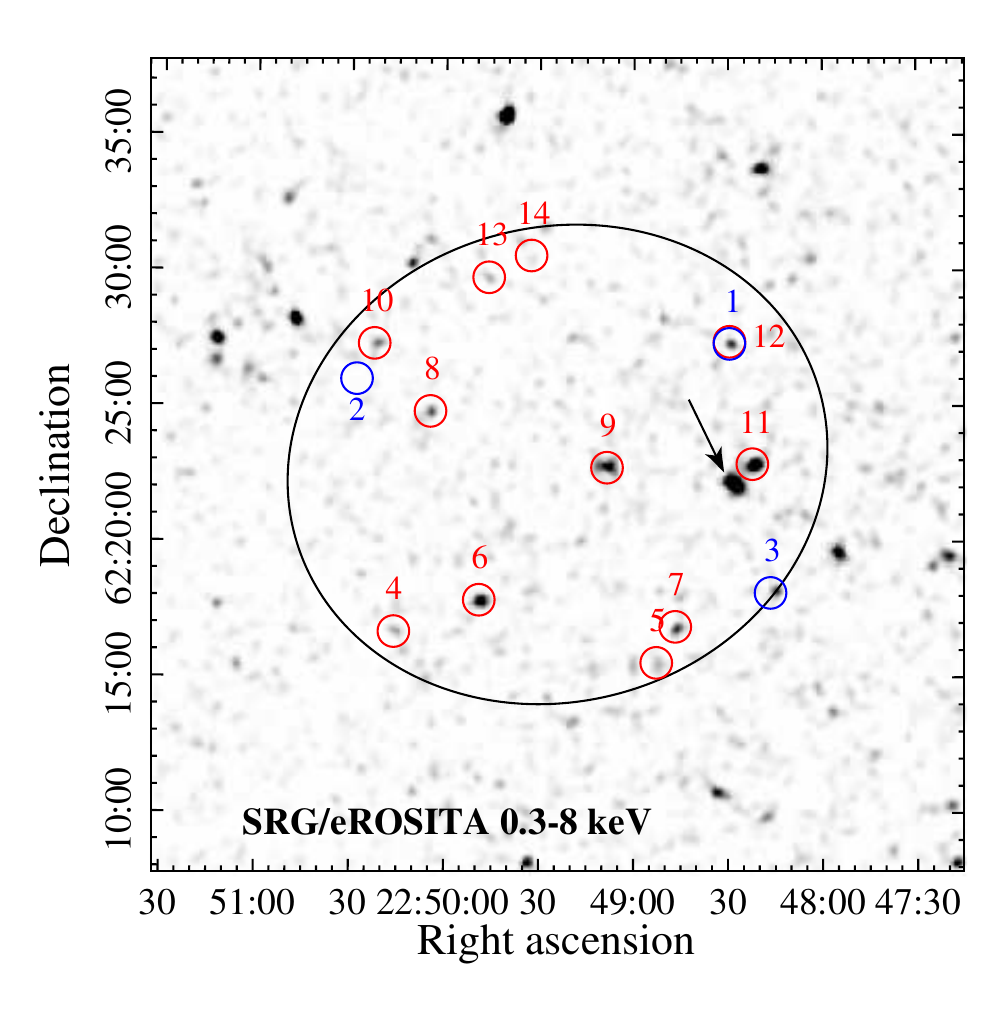}
\includegraphics[width=0.8\linewidth, trim={0.5cm 0.4cm 0.4cm 0.5cm}, clip]{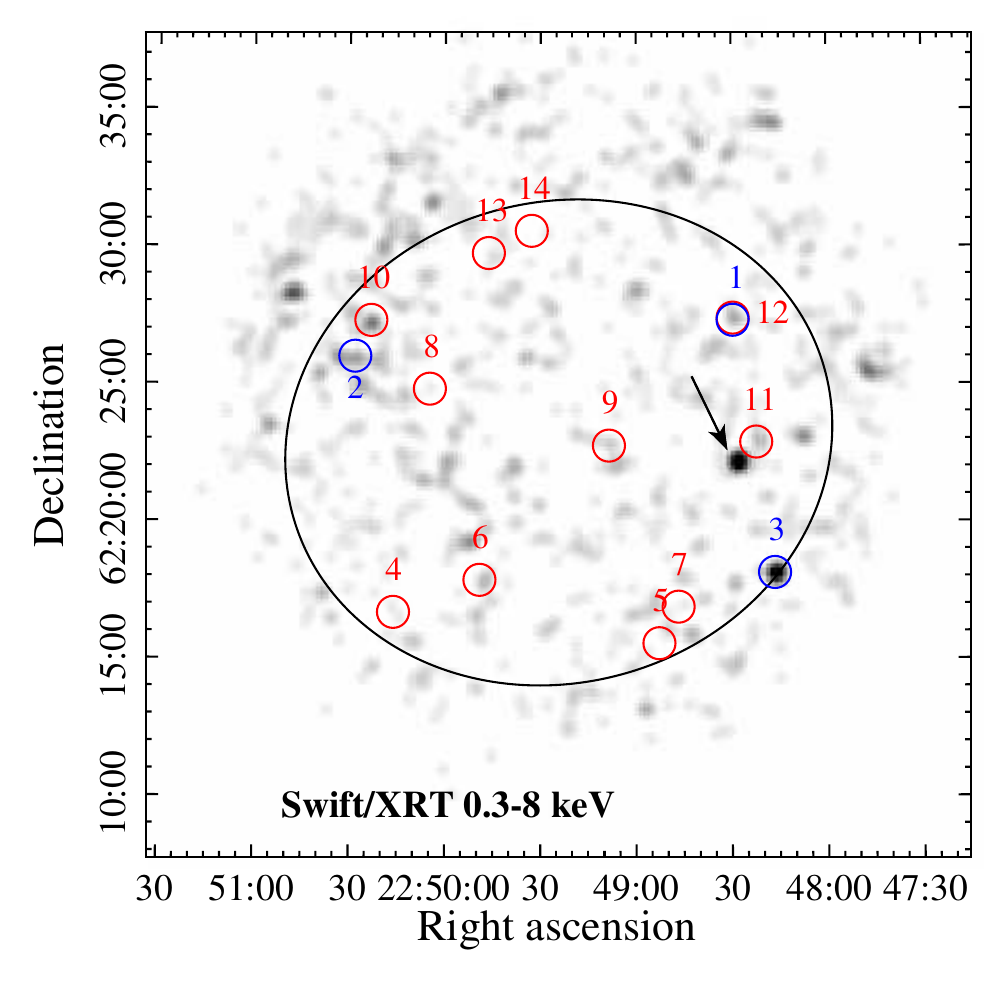}
\includegraphics[width=0.8\linewidth, trim={0.0cm 0.7cm 0.5cm 0.3cm}, clip]{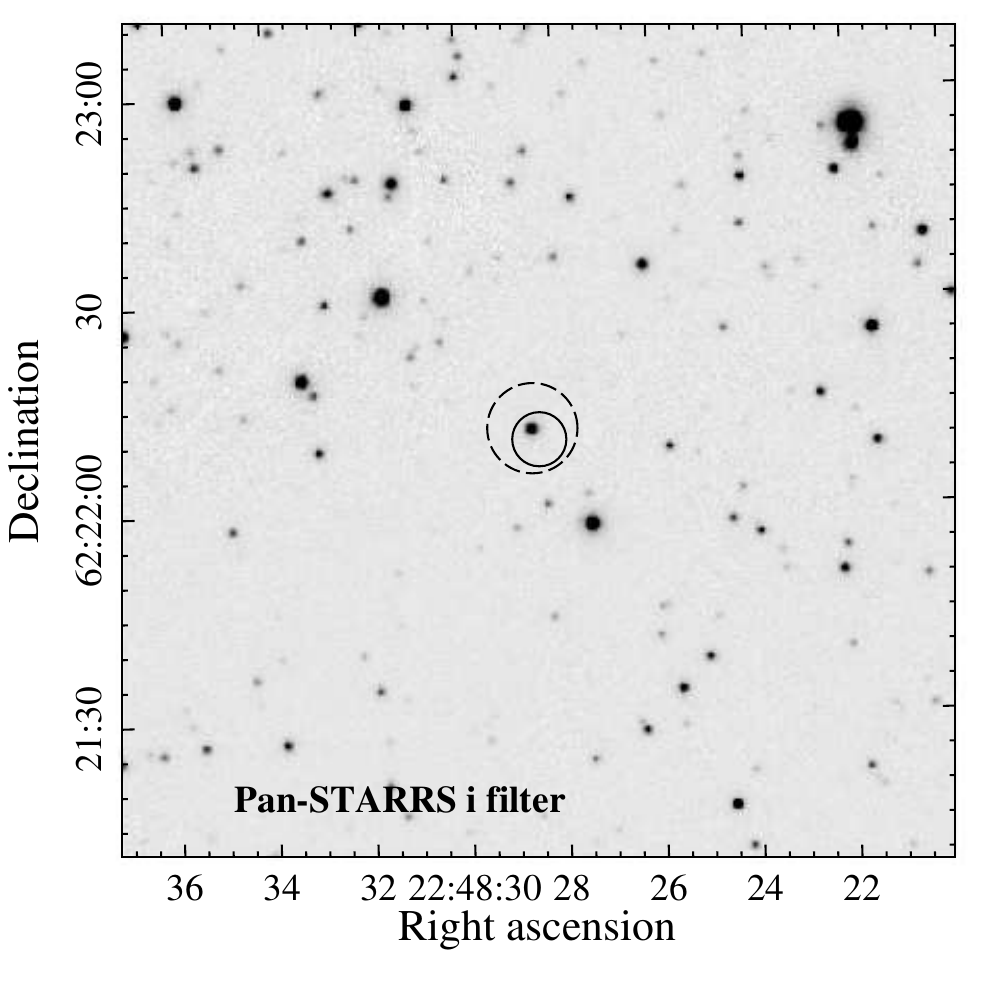}
\end{center}
\caption{$30 \times 30$~arcmin$^2$ \eros\ ({\it top panel}) and \sw\ ({\it middle panel}) images in the 0.3--8 keV range.  
The ellipse shows the 95\% position uncertainty of the  \src\ $\gamma$-ray position.
The likely X-ray counterpart of \src\ is marked by the arrow.
Other X-ray sources detected within the ellipse with \sw\ and \eros\ are shown by blue and red circles, respectively, and numbered.
{\it Bottom}: $2 \times 2$~arcmin$^2$ \ps\ image in the $i$ band. 
The solid and dashed circles show the 90\% position uncertainties of the X-ray source obtained with \eros\ and \sw, respectively. 
The likely optical counterpart is seen inside the circles.  }
\label{fig:img}
\end{figure}

To search for X-ray counterparts to \src, we investigated X-ray data obtained in the course of five all-sky surveys in 2020--2022 with the extended ROentgen Survey with an Imaging Telescope Array (\eros) telescope \citep{erosita2021} aboard the Spectrum-RG (SRG) orbital observatory \citep{Sunyaev2021}. 
The \src\ field as observed by \eros\ 
is shown in Fig.~\ref{fig:img}, top. 
The brightest source within the $\gamma$-ray positional uncertainty ellipse, \srge, is marked with the arrow.
Its coordinates are $\alpha_X(2000)$~=~22\h48\m28\fss43 and $\delta_X(2000)$~=~$+$62\degs22\amin10\farcs1 and 90 per cent position uncertainty is 3.9 arcsec.

We also checked the Living \sw-XRT Point Source (LSXPS) catalogue \citep{evans2023}.
At the \srge\ position there is the source \lsxps\ (Fig.~\ref{fig:img}, middle) with coordinates $\alpha_X({\rm J2000})$~=~22\h48\m28\fss57 and $\delta_X({\rm J2000})$~=~$+$62\degs22\amin11\farcs7 and 90 per cent position uncertainty 6.5 arcsec.

Then we examined the Gaia Data Release (DR) 3 catalogue \citep{gaia2016,gaia2023} and found a counterpart candidate to the X-ray source -- Gaia DR3 2207925451446739072 with the magnitude $G\approx18$ and effective temperature of about 3500 K. 
Its coordinates are $\alpha_{\rm opt}({\rm J2016})$~=~22\h48\m28\fss614 and $\delta_{\rm opt}({\rm J2016)}$~=~$+$62\degs22\amin11\farcs40 and  its proper motion is 12.3(1) \masyr. 
The geometric and photogeometric distances to the source are similar, $D_{\rm g}$~=~525--572 pc and $D_{\rm pg}$~=~537--596 pc \citep{bailer-johnes2021}. 
Thus, it definitely locates in the Galaxy.
The interstellar absorption in this direction for the object distance is $E(B-V)$ = 0.30--0.41 mag \citep{dustmap2019}.

The source is also presented in the Panoramic Survey Telescope and Rapid Response System survey (\ps) DR~2  catalogue \citep{ps2020} where its designation is PSO J342.1192$+$62.3698.
It remains the only counterpart candidate to \srge\ (see Fig.~\ref{fig:img}, bottom) even though the Pan-STARRS survey is deeper than the \gaia\ data.
In addition, the source was found in the Zwicky Transient Facility (ZTF) catalogue of periodic variable stars \citep{ztf-var} which is based on the DR 2 data covering approximately 1.3 yr. 
The source is listed as  ZTF~J224828.61$+$622211.4 and its period derived  
from the $r$-band light curve is 0.2334096 d. 
The $g$-band light curve gives a shorter period of 
0.2090120 d, though the low significance makes this estimate unreliable.


\section{Optical data}
\label{sec:lc}

\subsection{Optical light curves and the orbital period}

\begin{figure}
\begin{center}
\begin{minipage}{1\linewidth}
\includegraphics[width=0.95\linewidth,clip]{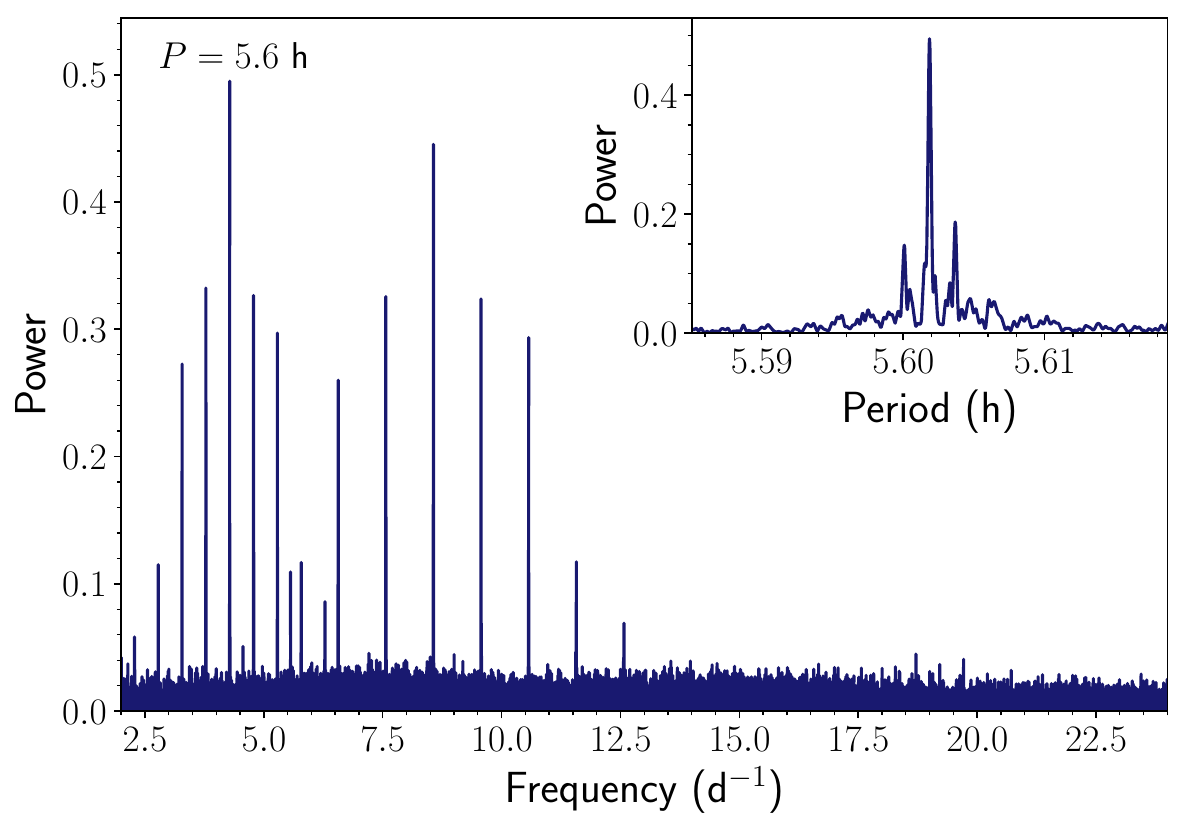}
\end{minipage}
\end{center}
\caption{Lomb-Scargle periodogram of the \src\ optical counterpart candidate obtained using the $r$-band ZTF data and two harmonics. 
The best period $P=5.6$~h corresponding to the highest peak is marked and the peak is enlarged in the
inset. }
\label{fig:ls}
\end{figure}

\begin{figure}
\begin{center}
\begin{minipage}{1\linewidth}
\includegraphics[width=0.95\linewidth,clip]{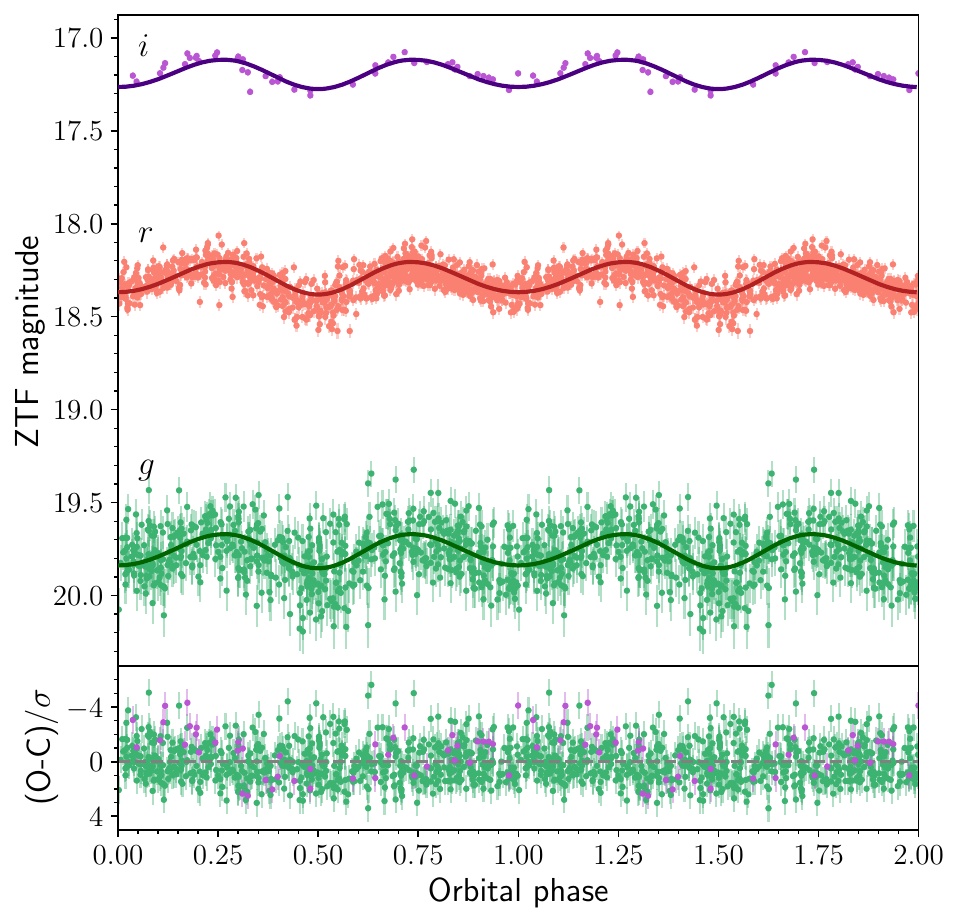}
\end{minipage}
\end{center}
\caption{\textit{Top}: ZTF light curves of the \src\ optical counterpart candidate in $g$, $r$ and $i$ bands folded with the presumed orbital period of 5.60187 h. 
Two periods are shown for clarity. 
The best-fitting model is shown by the solid lines. 
The $r$-band light curve was excluded from the fitting (see text).
The phase 0.0 is defined as the time when the secondary
is placed between the pulsar and an observer.
\textit{Bottom}: Residuals derived for each data point as the difference between the  observed (O) and the calculated (C) magnitudes divided by the error $\sigma$.}
\label{fig:ztf-lc}
\end{figure}

To check the period, we used the data from the ZTF DR 23 catalogue \citep{ztf2019} which covers $\approx$6.6 yr and contains about 1000 measurements in the $r$ band. 
We applied the {\sc nifty-ls} Lomb-Scargle periodogram python package \citep{nifty} to them using one and two harmonics.
The resulting power spectrum is shown in Fig.~\ref{fig:ls}.
The highest peak corresponds to the period\footnote{The period uncertainty was calculated as the half width at half maximum of the peak.} $P=0.2334112(57)\ {\rm d} = 5.60187(14)\ {\rm h}$.

We also checked the ZTF data in the $g$ band, which  contain
about 650 measurements. We obtained the same best period (5.60186(14) h) although the corresponding power is significantly lower, mainly  due to the high noise level of the data. 
Thus, the derived period is consistent with the value for the $r$-band data from the catalogue mentioned in Sec.~\ref{sec:data}. 

The ZTF light curves in $g$, $r$ and $i$ bands folded with the period~$P$ are shown in Fig.~\ref{fig:ztf-lc}.
They show two humps per the period and one minimum is slightly deeper than another.


\subsection{Light curves modelling}
\label{subsec:lc-fit}

\begin{table}
\renewcommand{\arraystretch}{1.2}
\caption{The light curve fitting results for \src.}
\label{tab:fit} 
\begin{center}
\begin{threeparttable}
\begin{tabular}{lc}
\hline
\multicolumn{2}{c}{Fitted parameters}                     \\
\hline
NS mass $M_{\rm NS}$, \msun                             & 1.53$_{-0.17}^{+0.05}$ \\
Mass ratio $q$ =  $M_{\rm c}/M_{\rm NS}$                & 0.35(3)\\ 
Distance $D$, pc                                       & 525(25)\\
Effective temperature $T_{\rm c}$, K                 & 3560(50) \\
Inclination $i$, deg                                    &  50(30)\\
Roche lobe filling factor $f_x$                         &1.00$_{-0.20}^{+0.00}$ \\
Reddening $E(B-V)$                                                  &0.35(5) \\
\hline
$\chi^2$/d.o.f.                                         & 740/672 \\
\hline
\multicolumn{2}{c}{Derived parameters}                   \\
\hline
Companion mass $M_{\rm c}$, \msun                       & 0.54 \\
Companion radius $R_{\rm c}^x$, \rsun                   & 0.80 \\
Companion radius $R_{\rm c}^y$, \rsun                   & 0.60 \\
\hline
\end{tabular}
\footnotesize{\textit{Notes.}
Numbers in parentheses denote 1$\sigma$ uncertainties related to the last significant digits. 
$R_{\rm c}^x$ and $R_{\rm c}^y$ are the radii of the ellipsoidal companion. The former is along the line passing through the centres of the binary sources.
d.o.f. $\equiv$ degrees of freedom.
}
\end{threeparttable}
\end{center}
\end{table}

To estimate the parameters of the presumed spider system, we fitted the folded light curves with the symmetric direct heating model. 
The details of the model can be found in \citet{zharikov2013, zharikov2019}.

\begin{figure*}
\begin{center}
\begin{minipage}{1\linewidth}
\includegraphics[width=1.0\linewidth,clip]{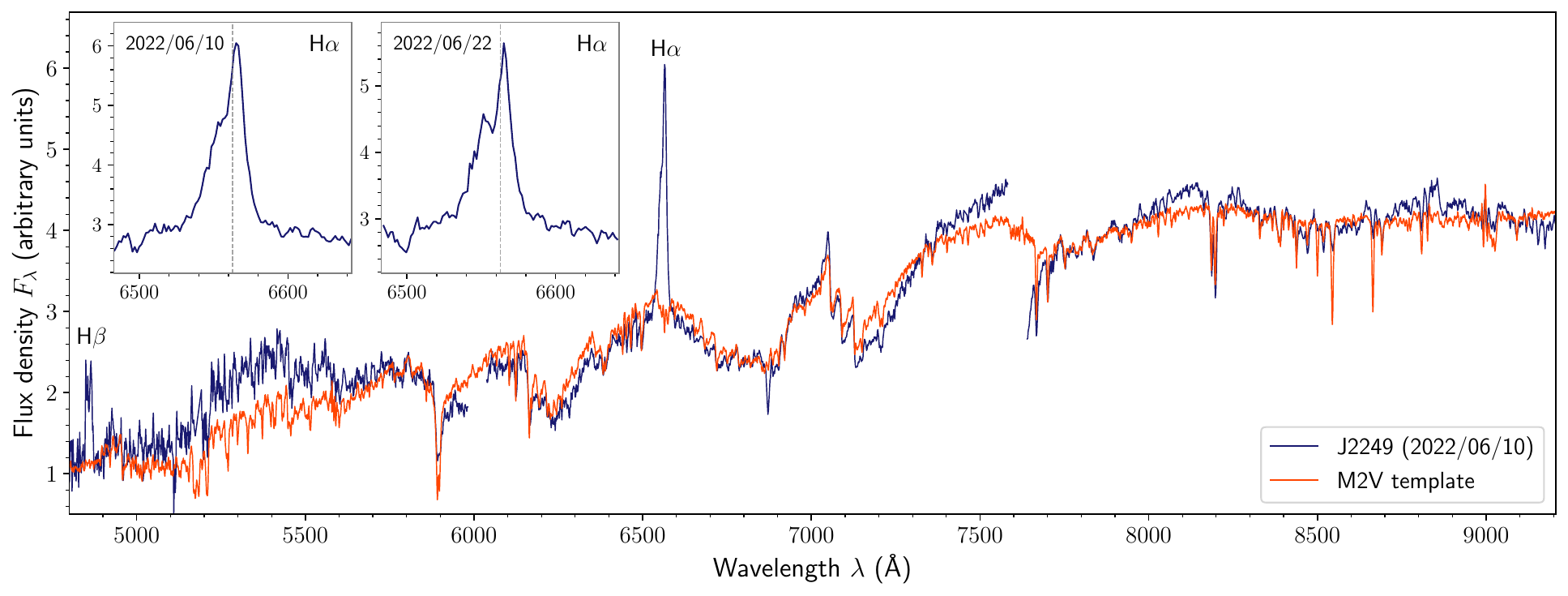}
\end{minipage}
\end{center}
\caption{Dereddened optical spectrum of the likely \src\ counterpart (dark blue) obtained with the \gem-North telescope on 2022 June 10  
(the empty regions represent instrumental chip gaps).
The orange line represents the M2-star template.
Insets show the zoomed-in region around the H$\alpha$ line from the spectra obtained on June 10 and 22.
The rest wavelength (6562.8 \AA) is marked by the dashed gray line. 
}
\label{fig:opt-spec}
\end{figure*}

The model parameters are the distance to the binary system $D$, the reddening $E(B-V)$, the pulsar mass $M_{\rm NS}$, the mass ratio of the binary components $q$, the system orbit inclination $i$, the effective irradiation factor $K_{\rm irr}$ defining the companion heating, the companion Roche lobe filling factor $f$ and the companion `night-side' temperature $T_{\rm n}$. 

We fixed the orbital period at $P_{\rm orb}=5.60187~{\rm h}$. 
The effective `night-side' temperature of the secondary, its irradiation, the Roche-lobe filling factor, the inclination, and the system mass ratio were initialized using a random sampling method.
The initial primary mass\footnote{The value serves as a starting guess for the optimization rather than a statistical prior.} 
was set to the canonical NS mass of 1.4~\msun, while initial values of the distance and interstellar extinction were randomly drawn from the ranges allowed by the \textit{Gaia} and dust map \citep{dustmap2019} estimates. 

The minimum of the $\chi^2$ function was determined using a gradient-descent algorithm, which was chosen because it substantially reduces the computational effort required for the minimization. The parameter uncertainties were calculated following the method proposed by \citet*{1976ApJ...208..177L}.

We found that the observed light curves can be reproduced without including irradiation of the secondary, i.e.\ $K_{\rm irr}=0$ and $T_{\rm n}\equiv T_{\rm c}$, where $T_{\rm c}$ is the effective temperature of the companion.  Furthermore, a pure blackbody approximation for the emission from the spider pulsar companions does not allow a simultaneous fit to all photometric bands.  This is largely because the spectral energy distribution of cool companions deviates significantly from a blackbody, requiring more sophisticated atmosphere models to account for molecular absorption and other non-trivial atmospheric effects (see for example \citealt{sullivan&romani2024}).
Nevertheless, the best-fitting model reproduces the $g$ and $i$ bands well, but  overestimates the flux in the $r$ band. Therefore, only the $g$ and $i$ bands together were used to determine the best-fitting parameters, and 
the $r$-band model was subsequently adjusted by a constant magnitude offset ($\Delta r = +0.26$ mag), representing the flux difference between a blackbody and an M-dwarf atmosphere at the derived effective temperature.

The results of the fit are presented in Table~\ref{tab:fit}, and the best-fitting models of the $g$-, $r$-, and $i$-band light curves are shown as solid lines in Fig.~\ref{fig:ztf-lc}. The reported $\chi^2$ is based on the $g$ and $i$ bands used for the minimization. Inclusion of the $r$-band data after the correction mentioned above results in $\chi^2$/d.o.f. = 1435/1674.
The upper limits on the primary mass and the mass ratio arise from the requirement that the Roche-lobe radius of the secondary must not be smaller than the radius of a zero-age main sequence star with a mass of $M_{\rm c} = q\,M_{\rm NS}$. 
For parameter values within these limits, the results are insensitive to the adopted initial conditions including starting mass of the pulsar.


\subsection{Archival optical spectroscopy with \gem}
\label{subsec:opt-spec}

The optical spectral observations\footnote{PI S. Swihart, programme ID GN-2022A-Q-236} of the \src\ counterpart candidate were performed with the \gem-North telescope on 2022 June 10 and 22. 
Two long-slit spectra with 12-minute exposures were obtained using the \gem\ Multi-Object Spectrograph with the R400+\_G5305 grating in conjunction with the GG455\_G0305 longpass filter covering 4500--9300 \AA. 
The slit width was 1\asec\ and the resulting resolution was 8 \AA\ at a blaze wavelength of 7640 \AA. 

Data reduction were performed using {\sc gemini} reduction software based on  
the Image Reduction Analysis Facility ({\sc iraf}) package. 
To calibrate the flux, the spectra of the spectrophotometric standard EG13  were used.

Note, that both spectra of the \src\ likely counterpart were obtained approximately at the same orbital phase, $\phi \sim 0.3$.
The resulting spectrum from June 10 is presented in Fig.~\ref{fig:opt-spec}. 
It was dereddened using the extinction law by \citet{Fitzpatrick1999} and $E(B-V)=0.35$ mag derived from the light curves fitting. The continuum of spectrum corresponds to an usual M-type dwarf star but there are Balmer emission (H$\alpha$ and H$\beta$) lines too.  The M2V-template\footnote{The template can be found at \url{https://svo2.cab.inta-csic.es/theory/newov/templates.php?model=tpl_kesseli}} from \citet{Kesseli2017} is also shown for comparison.  The source spectrum is generally consistent with the template, although we note that there is a slight flux excess in the $\sim$5000--5600 \AA\ range. 
The contribution of the excess in $g$-band light curve appears to be relatively weak and therefore not prominent in the rather noisy ZTF data.

The profiles of the H$\alpha$ emission line are shown in the insets of Fig.~\ref{fig:opt-spec}. The line is asymmetric, probably double-peaked, or has at least two distinct components.  
The red part of the profile is more intense than the blue one.
The line has an equivalent width of EW$_{\rm H\alpha}\approx -30~\text{\AA}$ and a full width at half maximum of 
$\mathrm{FWHM}_{\rm H\alpha}\approx 20~\text{\AA}$ (or $\sim$900 km s$^{-1}$). 
We also fitted the  
line profile with the double-gaussian model. 
For both spectra, we obtained the similar separation between the peak positions of the components of about 9 \AA\ (or $\sim$400 km~s$^{-1}$). 


\section{X-ray spectra}
\label{sec:x-rays}

\begin{figure}
\begin{center}
\begin{minipage}{1\linewidth}
\includegraphics[width=1.0\linewidth,clip]{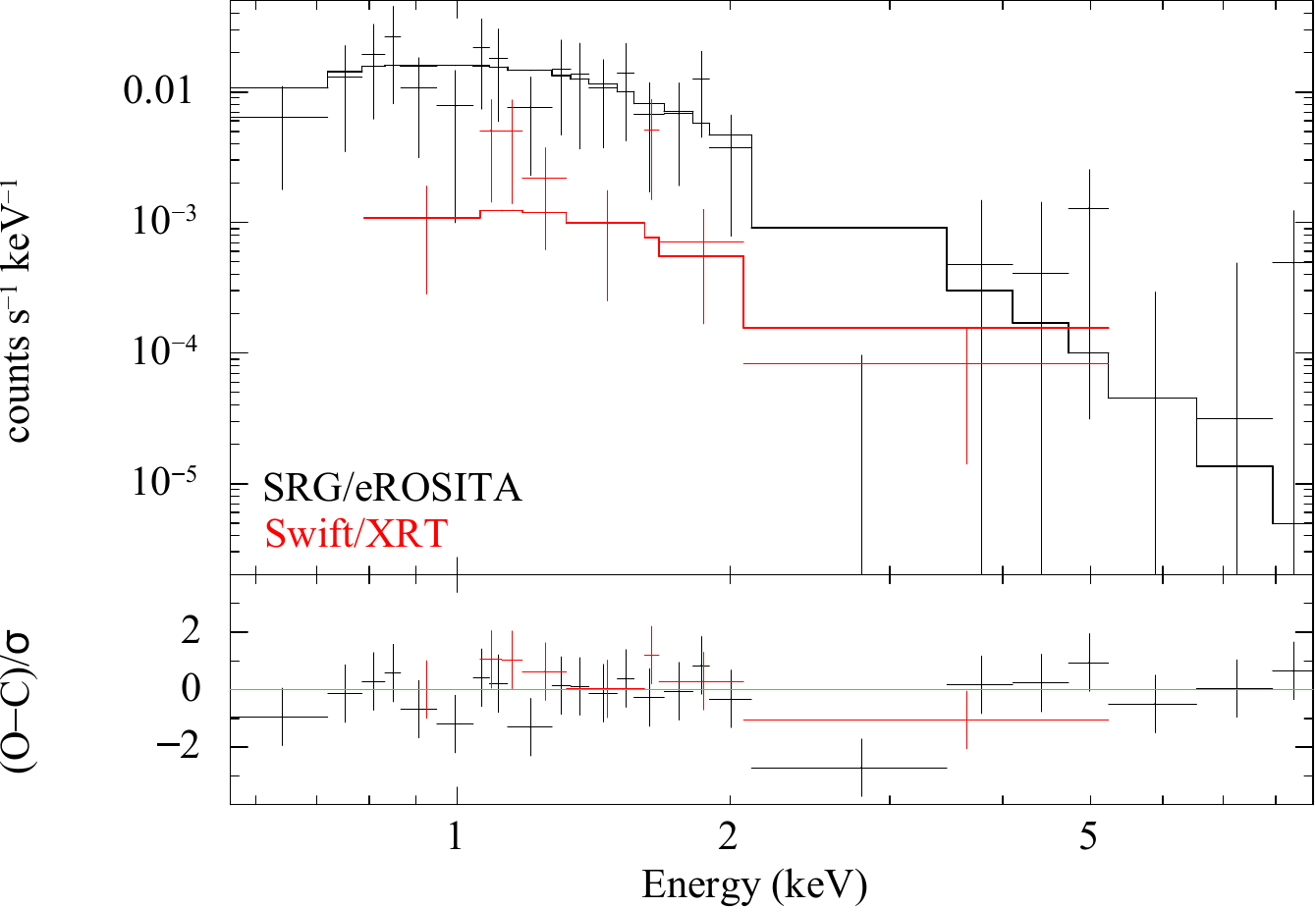}
\end{minipage}
\end{center}
\caption{\textit{Top}: the X-ray spectrum of the \src\ counterpart candidate with the best-fitting PL model.
The data obtained by different instruments are marked by different colours as indicated in the panel. 
For illustrative purposes, the \eros\ and \sw\ spectra were grouped to ensure at least 3 and 2 counts per energy bin, respectively.
\textit{Bottom}: residuals derived for each data point as the difference between the observed (O) and the calculated (C) flux density divided by the error $\sigma$.}
\label{fig:xray-spec}
\end{figure}

To study the X-ray spectrum of the \src\ counterpart candidate, we used SRG/\eros\ and archival \sw/XRT data.
From the \eros\ data the source spectra were extracted using a circular region with a radius of 60 arcsec centred at the source position. 
For the background extraction, we used an annulus region with the inner and outer radii of 150 and 300 arcsec around the source. 
The background sources were excluded with 40 arcsec radius apertures\footnote{Two sources were excluded from the background region and one (number 11 in Fig.~\ref{fig:img}) -- from the source region.}.
45.4 net counts were collected in the 0.3--9 keV band in the total exposure time of $\approx$2.7 ks (the vignetting corrected exposure is $\approx$1.4 ks).

The source was observed with \sw\ in 2019--2020 (ObsIDs 03110575001--03110575011) with the total exposure of $\approx$6.5 ks.
The \sw\ spectrum was extracted utilising the \sw-XRT data products generator\footnote{\url{https://www.swift.ac.uk/user_objects/}} \citep{evans2009}. 
This resulted in 15.7 net counts in the 0.3--10 keV band.

Both \eros\ and \sw\ spectra were grouped to ensure at least one
count per energy bin and fitted with the X-Ray Spectral Fitting Package ({\sc xspec}) v.12.15.0 \citep{xspec}.
The interstellar absorption was taken into account using the {\sc tbabs} model with the {\sc wilm} abundances \citep*{wilms2000}.

The reddening $E(B-V)=0.35$ obtained from the light curve fitting was transformed to the absorbing column density \nh~=~$3\times10^{21}$ cm$^{-2}$ utilising the empirical relation from \citet{foight2016}.
This value was fixed during the fitting procedure. 
The number of counts is low so we used the $C$-statistics \citep{cash1979}.

We applied the absorbed power-law (PL) model and found that the best-fitting parameters obtained for the \eros\ and \sw\ spectra are in agreement within their 1$\sigma$ uncertainties.
Thus, we fitted both spectra simultaneously which results in the photon index $\Gamma = 2.0^{+0.4}_{-0.3}$, the unabsorbed flux in the 0.5--10 keV band $F_X = (1.4 \pm 0.3) \times 10^{-13}$ \flux\ and $C / \text{d.o.f.} = 70/86$.
The spectra and the best-fitting model are presented in Fig.~\ref{fig:xray-spec}.


\section{Discussion}
\label{sec:discussion}

We have found a likely X-ray and optical counterpart to the unassociated \fermi\ source \src.
Its \gaia\ magnitude $G=18.01$ and colour $BP-RP=2.33$ combined with the 
observed X-ray flux in the 0.2--12 keV band of $\approx1.2\times10^{-13}$ \flux\ 
provides the X-ray to optical flux ratio of $\sim$0.2.
According to fig. 1 in \citet{rodriguez2024}, this implies that the source can belong to the spider pulsar family.

The optical light curves of the source are typical for RB systems where ellipsoidal modulations strongly dominate over the effect of the companion heating by the pulsar. 
Moreover, the irradiation effect here is small or even absent as observed for some RBs, e.g., PSR J1622$-$0315 or PSR J1816$+$4510 \citep{turchetta2023, Koljonen&Linares2023}.

The effective temperature of the \src\ counterpart candidate derived from the light curves fitting is 3560(50) K which is close to the M2V-type star. 
This is confirmed by the optical spectroscopy. The mentioned in Sec.~\ref{subsec:opt-spec} flux excess in the $\sim$5000--5600 \AA\ range may be attributed to a hot spot on the companion's surface as observed in some systems \citep[e.g.][]{swihart2019, kandel2020}. However, additional time-resolved spectroscopic and photometric observations are needed to confirm its presence.
Based on these results, \src\ can have one of the coldest companions among known RB systems. 
Indeed, typical temperatures are higher, about 4000--6000~K \citep[e.g.][]{turchetta2023}.
The even lower base (`night-side') temperature of $\approx$3300 K was obtained only for PSR J2339$-$0533 \citep{kandel2020}. 
In addition, PSR J1628$-$3205 can have a rather cold companion though its temperature is quite uncertain, 3560--4670 K \citep*{Li&Halpern2014}.

With a mass of $\approx$0.5~\msun\, the \src\ likely counterpart resides in the high-mass tail of the RB companion population \citep{strader2019}, possibly making it one of the most massive known in this class.
Other examples of massive sources are, e.g., companions of PSR J1306$-$4035 \citep[$M_{\rm c} = 0.51^{+0.02}_{-0.01}$~\msun;][]{swihart2019}, 
PSR J1803$-$6707 \citep[$M_{\rm c} = 0.44^{+0.05}_{-0.04}$~\msun;][]{Phosrisom2026}, tMSP candidate 3FGL J0427.9$-$6704 \citep[$M_{\rm c} = 0.65(8)$;][]{strader2016} and RB candidate  1FGL J0523.5$-$2529 \citep[$M_{\rm c} \gtrsim 0.8$~\msun;][]{strader2014}.

The most interesting feature in the optical spectrum of the \src\ likely counterpart is the Balmer emission lines. 
Such emission lines (sometimes clearly double-peaked) were observed for some spider pulsars.
For example, PSR J2339$-$0533 mentioned above shows strong H$\alpha$ and weaker H$\beta$ emission lines in some spectra \citep{kandel2020}.
Other examples are RBs PSR J1048$+$2339 \citep{MiravalZanon2021}, PSR J1306$-$4035 \citet{swihart2019}, PSR J1628$-$3205 \citep{strader2019} and PSR J0838$-$2827 \citep*{HalpernStraderLi2017}, RB candidate 1FGL J0523.5$-$2529 \citep*{HalpernPerezBogdanov2022}, BW PSR J1311$-$3430 \citep*{romani2015}, BW candidates 
4FGL J1408.6$-$2917 \citep{swihart2022}, ZTF J1406$+$1222 \citep{burdge2022} and 4FGL J0935.3$+$0901 \citep{wang2020,halpern2022}.
These sources also demonstrate flaring activity and variable heating.
The emission lines are assumed to be associated with the IBS 
or with the companion wind/chromosphere. 
For the \src\ counterpart candidate, the broadness of the H$\alpha$ line excludes its chromospheric origin (see 
\citealt{Mdwarf-Halpha}) and allows one to associate it with the IBS. 

The distance to the source obtained from the light curves modelling, 525(25) pc, agrees with the geometric and photogeometric values. 
For this distance, the proper motion $\mu = 12.3(1)$~\masyr\ corresponds to the transverse velocity $v_t$~=~29--32 km~s$^{-1}$ which is in compatible with velocities measured for binary pulsars \citep{hobbs2005}.
In addition, the best-fitting reddening agrees with the range, provided by the dust map of \citet{dustmap2019}, 0.30--0.41 mag.

The source's X-ray luminosity in the 0.5--10 keV band is $L_X$~=~(3.3--6.2)~$\times$~10$^{30}$ \ergs\ for 500--550 pc.
This value is in agreement with the luminosity distribution for RBs, although it lies at the lower bound of the latter: 
only one RB PSR J1816+4510 has similar low X-ray luminosity  \citep{Koljonen&Linares2023}.
The $\gamma$-ray luminosity is $L_\gamma$~=~(2.4--4.3)~$\times$~10$^{32}$ \ergs.
This is also rather low for spider pulsars which typically have luminosities $>$10$^{33}$ \ergs\ \citep{spidercat}.

The \src\ measured photon index $\Gamma = 2.0^{+0.3}_{-0.4}$ is consistent with the values observed for low-luminosity spider pulsars ($L_X \lesssim 10^{31}$ \ergs) while brighter RBs typically show harder non-thermal spectra \citep[e.g.][]{kandel2019, sullivan&romani2025}.
The softer emission in low-luminosity systems is likely attributed to a larger contribution from the heated polar caps relative to the IBS emission \citep{Koljonen&Linares2023}.

By its X-ray to optical flux ratio, the \src\ presumed counterpart also agrees with accreting systems such as LMXBs or cataclysmic variables (CVs). 
However, the source has an overall harder X-ray spectrum than LMXBs in the low-luminosity regime ($L_X \lesssim 10^{32}$--$10^{33}$ \ergs) when a thermal component from the bulk of the NS surface dominates \citep{tanaka1997}. 
As for CVs with appropriate periods and donors, the interpretation cannot be fully ruled out.
However, no strong flares or transition between low and high accretion states typical for many CV subclasses \citep{Inight2023} are seen in the \ps\ (MJD~$\approx$55000--57000) and ZTF (MJD~$\approx$58200--60600) data. 
The source is also stable in the \gaia\ data covering the time gap between \ps\ and ZTF.
The optical spectra show no evidence of the white dwarf emission (though it could be very cold) or bright accretion disc/flow suggesting a very low accretion rate.
We could assume a low-accretion rate polar \citep[e.g.][]{Schwope2025A} but if so, the optical spectrum should exhibit pronounced cyclotron humps.
Thus, if the source is indeed a CV, it represents a highly unique object with a low accretion rate that has remained stable for at least $\sim$15 years.
Moreover, it cannot be associated with the \fermi\ source. 

We also investigated other X-ray sources marked in the top and middle panels of Fig.~\ref{fig:img} as well as their possible optical counterparts. 
The details are presented in Appendix~\ref{app:srcs}.
None of them seems to be responsible for the $\gamma$-ray emission.


\section{Conclusions}
\label{sec:conclusions}

To sum up, J2249 is likely a rare member of the RB family -- one with a cool companion
and unusually low X-ray and $\gamma$-ray luminosities.  Such systems are challenging 
to detect due to the limited depth of current optical and X-ray surveys. 
At a distance just three times greater ($\gtrsim$ 1.5 kpc), J2249 would have remained 
undetected by \eros\ and \sw, and its optical periodicity would have been missed in 
ZTF data.

Searching for pulsations in the radio from an NS are needed to confirm the RB nature of the \src\ likely optical/X-ray counterpart.
The present analysis is limited by the relatively noisy ZTF data and sparse spectroscopic coverage. Deeper, phase-resolved optical photometry and spectroscopy are therefore necessary to confirm and robustly constrain the system parameters, including the mass ratio and pulsar mass derived from radial velocity curves, and to investigate the IBS geometry.


\section*{Acknowledgements}

We thank the anonymous referee for useful and constructive comments.
This work has made use of data from the European Space Agency (ESA) mission {\it Gaia} (\url{https://www.cosmos.esa.int/gaia}), processed by the {\it Gaia} Data Processing and Analysis Consortium (DPAC, \url{https://www.cosmos.esa.int/web/gaia/dpac/consortium}). Funding for the DPAC has been provided by national institutions, in particular the institutions participating in the {\it Gaia} Multilateral Agreement.
The Pan-STARRS1 Surveys (PS1) and the PS1 public science archive have been made possible through contributions by the Institute for Astronomy, the University of Hawaii, the Pan-STARRS Project Office, the Max-Planck Society and its participating institutes, the Max Planck Institute for Astronomy, Heidelberg and the Max Planck Institute for Extraterrestrial Physics, Garching, The Johns Hopkins University, Durham University, the University of Edinburgh, the Queen's University Belfast, the Harvard-Smithsonian Center for Astrophysics, the Las Cumbres Observatory Global Telescope Network Incorporated, the National Central University of Taiwan, the Space Telescope Science Institute, the National Aeronautics and Space Administration under Grant No. NNX08AR22G issued through the Planetary Science Division of the NASA Science Mission Directorate, the National Science Foundation Grant No. AST-1238877, the University of Maryland, Eotvos Lorand University (ELTE), the Los Alamos National Laboratory, and the Gordon and Betty Moore Foundation.
Based on observations obtained with the Samuel Oschin Telescope 48-inch and the 60-inch Telescope at the Palomar Observatory as part of the \textit{Zwicky} Transient Facility project. ZTF is supported by the National Science Foundation under Grants No. AST-1440341 and AST-2034437 and a collaboration including current partners Caltech, IPAC, the Oskar Klein Center at Stockholm University, the University of Maryland, University of California, Berkeley , the University of Wisconsin at Milwaukee, University of Warwick, Ruhr University, Cornell University, Northwestern University and Drexel University. Operations are conducted by COO, IPAC, and UW.
This work made use of data supplied by the UK Swift Science Data Centre at the University of Leicester.
This work used data obtained with eROSITA telescope onboard SRG observatory. The SRG observatory was built by Roskosmos in the interests of the Russian Academy of Sciences represented by its Space Research Institute (IKI) in the framework of the Russian Federal Space Program, with the participation of the Deutsches Zentrum für Luft- und Raumfahrt (DLR). The SRG/eROSITA X-ray telescope was built by a consortium of German Institutes led by MPE, and supported by DLR.  The SRG spacecraft was designed, built, launched and is operated by the Lavochkin Association and its subcontractors. The science data are downlinked via the Deep Space Network Antennae in Bear Lakes, Ussurijsk, and Baykonur, funded by Roskosmos. The eROSITA data used in this work were processed using the eSASS software system developed by the German eROSITA consortium and proprietary data reduction and analysis software developed by the Russian eROSITA Consortium.
The work of AVK and DAZ (periodicity search, reduction of the \gem\ data, investigation of X-ray sources in the \fermi\ ellipse) was supported by the baseline project FFUG-2024-0002 of the Ioffe Institute. The analysis of the J2249 X-ray spectra by AVK was supported by the Russian Science Foundation project 22-12-00048-P.
DAZ thanks Pirinem School of Theoretical Physics for hospitality.
SVZ acknowledges DGAPA-PAPIIT grant  IN105826.
%

\section*{Data Availability}

The ZTF data are available through the archive \url{https://irsa.ipac.caltech.edu/Missions/ztf.html}, \sw/XRT data -- \url{https://www.swift.ac.uk} and eROSITA data -- on request.



\bibliographystyle{mnras}
\bibliography{ref} 



\appendix

\section{Other X-ray sources in the \fgl\ position uncertainty ellipse and their possible optical counterparts}
\label{app:srcs}

\begin{table*}
\caption{X-ray sources in the \fgl\ position uncertainty ellipse and their likely optical counterparts.}
\label{tab:srcs} 
\begin{center}
\begin{threeparttable}
\begin{tabular}{ccccl}
\hline
Err$_{90}$ & $\alpha_{\rm opt}$, $\delta_{\rm opt}$ & $f_X / f_{\rm opt}$ & $BP-RP$ & Comments  \\
\hline
\multicolumn{5}{c}{\textbf{1) LSXPS J224829.5+622718 / 12) SRGe J224829.6+622723}} \\
\hline 
7.8  &    22\h48\m28\fss69, +62\degs27\amin25\farcs2 & 0.03 & 1.95 & F-type star \\
\hline
\multicolumn{5}{c}{\textbf{2) LSXPS J225028.2+622559 -- variable in X-rays}} \\
\hline     
6.4  & A) 22\h50\m27\fss99, +62\degs26\amin00\farcs7 & 0.2  & 3.30 & late-type star, likely counterpart to the X-ray source \\
     & B) 22\h50\m28\fss19, +62\degs26\amin02\farcs5 & 0.8  & 2.15 & late-type star  \\
\hline
\multicolumn{5}{c}{\textbf{3) LSXPS J224816.5+6218 -- variable in X-rays}} \\
\hline
5.8  & 22\h48\m16\fss20, +62\degs18\amin09\farcs2 & 0.08 & 2.57 & RS CVn, $P_b\approx2.3$ d \citep{ztf-var} \\ \hline
\multicolumn{5}{c}{\textbf{4) SRGe J225016.1+621641}} \\
\hline
7.8  &    22\h50\m15\fss72, +62\degs16\amin47\farcs0 & 0.1  & 3.34 & late-type star \\
\hline
\multicolumn{5}{c}{\textbf{5) SRGe J224852.8+621533}} \\
\hline
9.7   &  22\h48\m53\fss24, +62\degs15\amin28\farcs2 & 0.02 & 3.35 & YSO \citep{gaia2023} \\ \hline
\multicolumn{5}{c}{\textbf{6) SRGe J224949.1+621751}} \\
\hline
7.3  &    22\h49\m48\fss55, +62\degs17\amin49\farcs6 & 0.006& 2.34 & active F-type star\\
\hline
\multicolumn{5}{c}{\textbf{7) SRGe J224846.8+621653}} \\
\hline
6.9  &    22\h48\m46\fss01, +62\degs16\amin54\farcs8 & 0.08 & 3.23 & late-type star, likely YSO \\
\hline
\multicolumn{5}{c}{\textbf{8) SRGe J225004.8+622448}} \\
\hline
7.1  &    22\h50\m04\fss76, +62\degs24\amin54\farcs5 & 0.002& 1.71 & active G-type star, likely binary\\
\hline
\multicolumn{5}{c}{\textbf{9) SRGe J224908.5+622244}} \\
\hline
7.4  &  22\h49\m08\fss56, +62\degs22\amin45\farcs8 & 0.01 & 2.42 & RS CVn, $P_b\approx1.8$ d \citep{ztf-var}
\\
\hline
\multicolumn{5}{c}{\textbf{10) SRGe J225022.7+622718}} \\
\hline
9.7  &  22\h50\m22\fss54, +62\degs27\amin20\farcs9 & 0.006& 2.38 & YSO \citep{gaia2023} \\
\hline
\multicolumn{5}{c}{\textbf{11) SRGe J224822.3+622253}} \\
\hline
5.2  & 22\h48\m21\fss82, +62\degs22\amin54\farcs3 & 0.006& 1.90 & active late-type star \\
\hline
\multicolumn{5}{c}{\textbf{13) SRGe J224946.3+622945}} \\
\hline
8.9  & 22\h49\m46\fss82, +62\degs29\amin48\farcs9 & 0.03 & 3.54 & YSO \citep{gaia2023}
\\
\hline
\multicolumn{5}{c}{\textbf{14) SRGe J224932.8+623033}} \\
\hline
8.7   & 22\h49\m32\fss11, +62\degs30\amin30\farcs5 & 0.03 & 3.22 & YSO \citep{gaia2023}
     \\ 
\hline
\end{tabular}
\footnotesize{\textit{Notes.} 
Err$_{90}$ is the 90 per cent position uncertainty of the X-ray source measured in arcsec; if the source is presented in both \sw\ and \eros\ data, the more precise position was used.  
$\alpha_{\rm opt}$ and $\delta_{\rm opt}$ are coordinates of the possible \gaia\ counterparts. 
$f_X$ is the observed X-ray flux in the 0.2--12 keV band.
$f_{\rm opt}$ is the observed optical flux calculated using the $G$-band magnitude.
RS CVn $\equiv$ RS Canum Venaticorum-type binary system, YSO $\equiv$ young stellar object.}
\end{threeparttable}
\end{center}
\end{table*}

Table~\ref{tab:srcs} presents X-ray sources found inside the \fgl\ 95 per cent position uncertainty ellipse, with the exception of the RB candidate.
They are numbered as in the top and middle panels of Fig.~\ref{fig:img}.
We note, that sources 2 and 3 were not detected with \eros\ that indicates their rather strong variability in X-rays, possibly flaring activity.
Other sources are too weak to make any definitive conclusions about their variability.

To clarify the nature of the sources, we also searched for their likely optical counterparts in the \gaia\ catalogue and estimated X-ray to optical flux ratios $f_X/f_{\rm opt}$.
Observed X-ray fluxes were calculated from the mean count rates using the WebPIMMS tool\footnote{\url{https://heasarc.gsfc.nasa.gov/cgi-bin/Tools/w3pimms/w3pimms.pl}} and assuming the absorbed PL model with the column density $N_{\rm H} = 3\times 10^{20}$ cm$^{-2}$ and photon index $\Gamma = 2$. 
Observed optical fluxes were obtained from the \gaia\ magnitude $G$.
If several optical sources coincide with the X-ray source, $f_X$/$f_{\rm opt}$ should be considered as an upper limit.

We also used information from the Pan-STARRS and ZTF catalogues as well as optical-infrared spectral energy distributions (SEDs) of the sources\footnote{SEDs were obtained through the VizieR Photometry viewer \url{https://vizier.cds.unistra.fr/vizier/sed/}} and distance estimates from \citet{bailer-johnes2021} to establish the most probable nature of the sources, which is mention in Table~\ref{tab:srcs}. 
Note, that we present only the most plausible optical counterpart for each X-ray source. 
If no compelling candidate is found, all optical sources within the X-ray positional error circle are reported.

\bsp	
\label{lastpage}
\end{document}